\documentclass[conference]{IEEEtran}
\IEEEoverridecommandlockouts
\usepackage{cite}
\usepackage{amssymb,amsfonts,amsmath}
\usepackage{algorithmic}
\usepackage{graphicx}
\usepackage{textcomp}
\usepackage{mathrsfs}
\usepackage{array, color,makecell}
\usepackage{setspace}

\def\BibTeX{{\rm B\kern-.05em{\sc i\kern-.025em b}\kern-.08em
    T\kern-.1667em\lower.7ex\hbox{E}\kern-.125emX}}
\begin{document}

\title{Robust Beamforming for SWIPT System with Chance Constraints\\
\thanks{This work is supported by National Natural Science Foundation of China under Grant No. 61771072 and No. 61427801, the State Major Science and Technology Special Projects of China under
Grant 2016ZX03001017-004. 
}
}

\author{\IEEEauthorblockN{Yinglei Teng, Wanxin Zhao, Mei Yan, Yong Zhang, Mei Song}
\IEEEauthorblockA{\textit{Beijing Key Laboratory of Space ground Interconnection and Convergence} \\
\textit{Beijing University of Posts and Telecommunications (BUPT)}\\
Beijing, China \\
{Email:lilytengtt@gmail.com,\{zhaowx,yanmei,yongzhang,songm\}@bupt.edu.cn}}
}

\maketitle

\begin{abstract}
The robust beamforming problem in multiple-input single-output (MISO) downlink networks of simultaneous wireless information and power transfer (SWIPT) is studied in this paper. Adopting the time switching fashion to perform energy harvesting and information decoding respectively, we aim at maximizing the sum rate under imperfect channel state information (CSI) and the chance constraints of users' harvested energy. In view of the fact that the constraints for minimal harvested energy is not necessary to meet from time to time, this paper adopts chance constraint to model it and uses the Bernstein inequality to transform it into deterministic constraints equivalently. Recognizing the maximum sum rate problem of imperfect CSI as nonconvex problem, we transform it into finding the expectation of minimum mean square error (MMSE) equivalently in this paper, and an alternative optimization (AO) algorithm is proposed to decompose the
optimization problem into two sub-problems: the transmit
beamformer design and the division of switching time. The simulation results show the performance gains compared to non-robust state
of the art schemes.
\end{abstract}


\section{Introduction}
In order to meet the explosive growth of data traffic in 5G wireless networks, the network density and coverage are increasing. Along with it, the energy problem has become the focus of attention. Considering that some nodes may not connect to the power grid directly, energy harvesting technology has attracted more and more attentions from academia to industry. Radio frequency (RF) signals can not only transmit information, but also be suitable for far field energy transmission, so it's a very promising way to transmit signals and energy for the nodes with no grid energy supply. Moreover, simultaneous wireless information and power transfer  (SWIPT) is considered as one of the promising technologies for the sustainable development of green communications. 
\par
Concerning SWIPT, there are mainly two basic patterns for the receiver to divide the received signal for information decoding (ID) and energy harvesting (EH), namely power splitting (PS) and time switching (TS)\cite{1-Ding2015Application}. In the PS pattern, information and energy are transmitted by the base station (BS) simultaneously using the same signal. At receiver, the received signal is divided into two streams by a power divider for ID and EH respectively. In the receive TS method, the information and energy are transmitted using the same signal, allowing users to decode the information in a fraction of the time and harvest energy in the rest of the time. In the transmit TS method, information and energy are transmitted by BSs at different time periods, and the user processes the received signals for ID and EH synchronously in time. Ali A. Nasir, et al compared these three methods in \cite{2-Nasir2017Beamforming}, and the results show that the receive PS is better than the receive TS, and the transmit TS is better than the receive PS, so this paper will carry on the SWIPT study on the TS fashion.

Taking advantages of the spatial diversity, transmission beamforming technology can improve energy efficiency of the system, and there have been notable results in the literature. \cite{3-Zong2016Optimal} studied the SWIPT in the MIMO interference channel of K-users, and the receivers use PS technology to divide the received signal into two streams for ID and EH respectively. The purpose of this paper is to maximize sum rate of the system through joint design of transmit beamformers under the constraints of power consumption, energy consumption and the energy collection. Finally, a semidefinite relaxation (SDR) based method is proposed to solve the nonconvex optimization problem. \cite{4-Chen2013Energy}\cite{5-Chen2014Wireless} studied the receive TS, where information and energy are transmitted using the same signal. These studies all assume the perfect channel state information (CSI), but it is difficult to obtain accurate CSI in the actual communication environment, so it is necessary to consider the existence of errors. In \cite{6-Le2017Robust}\cite{7-Chu2016Simultaneous}, the robust beamforming problem of SWIPT with the presence of CSI errors was investigated. In \cite{6-Le2017Robust}, the total transmission power consumption was minimized under the condition of guaranteeing data transmission reliability, data transmission security and energy transmission reliability, and the safe convex approximation is used to deal with the chance constraint problem. \cite{7-Chu2016Simultaneous} studied the maximum security rate of transmission signals with artificial interference, and proposed two level optimization algorithm and continuous convex approximation algorithm to solve the nonconvex optimization problems.

Regarding the existing studies, no one has yet considered CSI errors in the SWIPT scene with the transmit TS. In this paper, we will focus on SWIPT in the multiple-input single-output (MISO) network scenario with CSI errors. The users' sum rate is maximized under the condition of BS's total power constraint and users' harvested energy constraint. This is a typical nonconvex and stochastic optimization problem. In this paper, we will use minimum mean square error (MMSE) method to transform the maximum sum rate problem equivalently owing to they have the same optimal solution. 

The rest of the paper is organized as follows. Section 2 describes the network model and the basic problem. Section 3 formulates the maximization sum rate problem, and introduces the AO algorithm and WMMSE method to transform the original problem equivalently. Section 4 provides simulation results to compare the performance of all solutions. Finally, Section VI concludes the paper.

\section{System model and problem formulation}

\subsection{Transmission model and imperfect CSI model}

Consider the downlink channel of MISO networks which consists of \emph{L} BSs and \emph{K} users. Assume that each BS is equipped with ${N_l}$  transmit antennas while each user only has a single antenna. Employing the TS fashion, for a given time slot, a fraction of time $\tau $ is used for energy reception while the remaining fraction of time $1- \tau $ for information transfer. In this paper, it is assumed that time synchronization has been perfectly established between the transmitters and the receivers. The SWIPT system and TS protocol are shown in Fig.1.
\begin{center}
\includegraphics[width=2.2in]{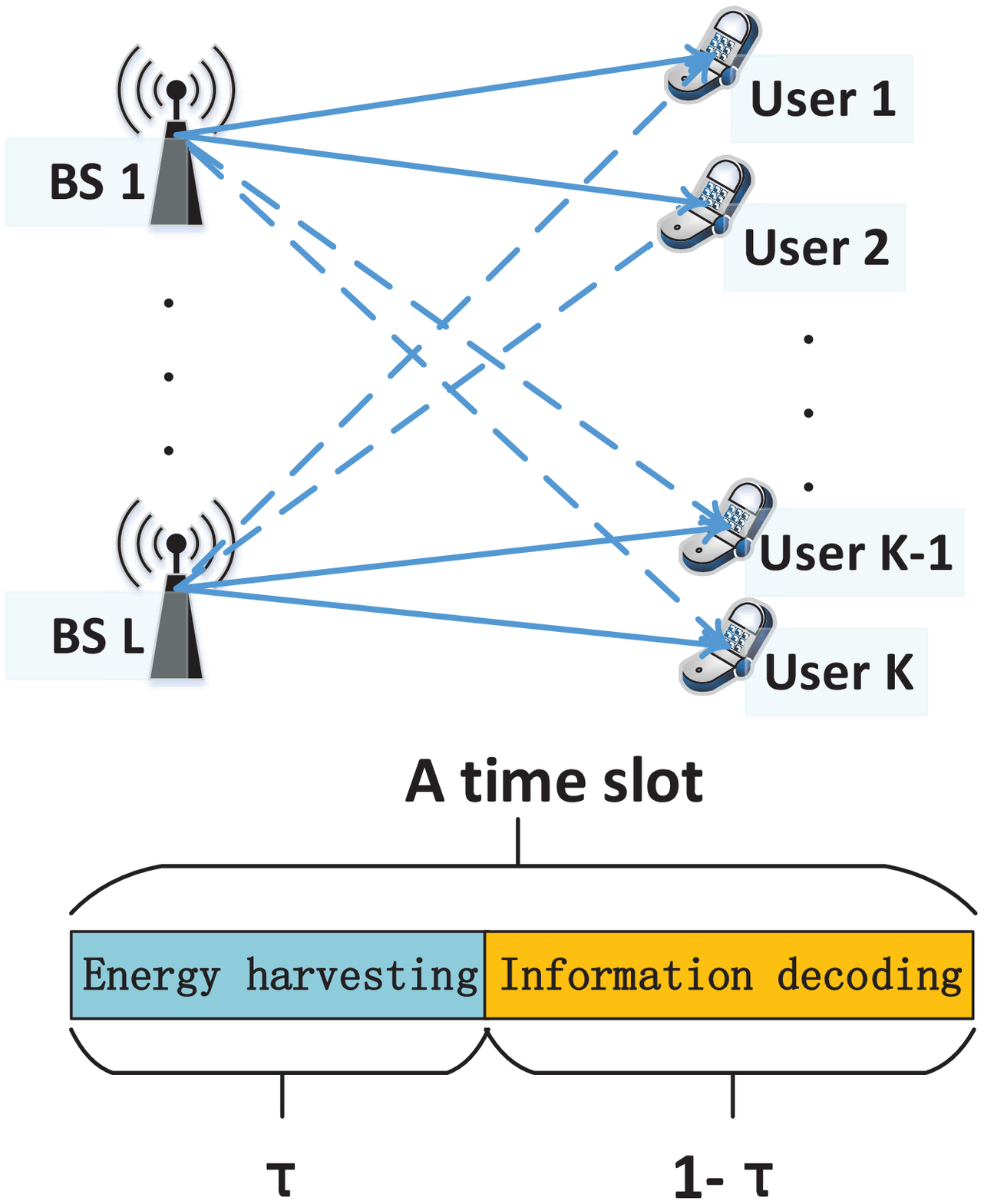}
\\
\footnotesize
Fig.1 Model of the SWIPT
\end{center}

Let ${\mathbf{v}_{kl}}$ denotes the beamforming vector from BS \emph{l} to user $k \in {\cal K} \buildrel \Delta \over = \left\{ {1, \cdots ,K} \right\}$ and ${\mathbf{v}_k} \buildrel \Delta \over = {[\mathbf{v}_{k1}^T,\mathbf{v}_{k2}^T, \cdots ,\mathbf{v}_{kL}^T]^T} \in {C^N}$ denotes the network-wide beamforming vector. Let ${\left(  \cdot  \right)^T}$ denotes the transpose operator, and ${\left(  \cdot  \right)^H}$ denotes the complex conjugate transpose
operator. Adopting the beamformer $\mathbf{v}_k^I,\mathbf{v}_{kl}^I$ and $\mathbf{v}_k^E,\mathbf{v}_{kl}^E$ for information transmission and energy transmission separately, the received signal at user \emph{k} is given by
\begin{equation}
{y_k^{\{ I,E\} }} = \mathbf{h}_k^H\mathbf{v}_k^{\{ I,E\} }{s_k} + \sum\limits_{j \ne k,j \in \mathcal{K}} {\mathbf{h}_k^H\mathbf{v}_j^{\{ I,E\} }{s_j}}  + {n_k},
\end{equation}
where ${\mathbf{h}_k} = {\left[ {\mathbf{h}_{k,1}^H,\mathbf{h}_{k,2}^H, \cdots ,\mathbf{h}_{k,L}^H} \right]^H} \in {C^{N \times 1}}$ denotes the channel between node \emph{k} and all BSs, and ${s_k} \sim {\cal C}{\cal N}(0,1)$ denotes the transmitted data symbol for user \emph{k}, $N = \sum\nolimits_{l = 1}^L {{N_l}}$ is the antenna number of all BSs, and ${n_k} \sim {\cal C}{\cal N}\left( {0,\sigma _k^2} \right)$ is the additive Gaussian noise.

It is assumed that only imperfect CSI is available in this paper. Specifically, the true channel vector ${\mathbf{h}_k}$ can be decomposed into
\begin{equation}
{\mathbf{h}_k} = {\widehat{ \mathbf{h}}_k} + {\mathbf{e}_k},
\end{equation}
where ${\widehat{ \mathbf{h}}_k}$ and ${\mathbf{e}_k}$ denote the channel estimate at transmitters and the CSI error respectively. Furthermore, ${\mathbf{e}_k}$ is complex Gaussian random vector with zero mean and covariance matrix ${\varepsilon _k} \succeq 0$, i.e. ${\mathbf{e}_k} \in {\cal C}{\cal N}(0,{\varepsilon _k})$. Therefore, the corresponding SINR of user \emph{k} can be expressed as
\begin{equation}
SIN{R_k} = \frac{{{{\left| {{{\left( {{{\widehat{ \mathbf{h}}}_k} + {\mathbf{e}_k}} \right)}^H}\mathbf{v}_k^I} \right|}^2}}}{{\sum\limits_{j \ne k,j \in \mathcal{K}} {{{\left| {{{\left( {{{\widehat{ \mathbf{h}}}_k} + {\mathbf{e}_k}} \right)}^H}\mathbf{v}_j^I} \right|}^2}}  + \sigma _k^2}}.
\end{equation}
\par
For information transfer, the achievable date rate during one transmission frame is given by 
\begin{equation}
{R_k} = (1 - \tau )\log \left( {1 + SIN{R_k}} \right),
\end{equation}
where $\tau $ is the TS ratio, the greater $\tau $ implies the greater proportion of time will be taken to harvest energy, and vice versa.

\subsection{Transmit Time Switching}
Unlike the power-splitting system, in this paper, a fraction time $0 < \tau  < 1$ is used for energy harvesting while the remaining time $1 - \tau $ for information decoding. The energy harvested by user \emph{k} is denoted as
\begin{equation}
{E_k} = \tau {\eta _k}\left( {\sum\limits_{j \in \mathcal{K}} {{{\left| {\mathbf{h}_k^H\mathbf{v}_j^E} \right|}^2}}  + {\sigma _k^2}} \right),
\end{equation}
where ${\eta _k}$ is the energy conversion efficiency. Because of the error of CSI, this paper uses the chance constraint method to guarantee the energy harvested by the user, and the chance constraints for minimal harvested energy are given as follows
\begin{equation}
{\rm{Pr}}\left\{ {{E_k} \ge E_k^{\min }} \right\} \ge 1 - \theta ,\forall k \in {\cal K},
\end{equation}
where $E_k^{\min }$ represents the required minimum energy of UE. If the receiver fails to meet this requirement, a energy outage occurs. $\theta  \in (0,1]$ indicates the maximum tolerable energy outage probability for user \emph{k}, and the smaller $\theta $ implies strictness for the harvested energy requirement. In addition, the power constraint of each BS is given by
\begin{equation}
\sum\limits_{k \in \mathcal{K}} {\tau \left\| {\mathbf{v}_{kl}^E} \right\|_2^2}  + \sum\limits_{k \in \mathcal{K}} {\left( {1 - \tau } \right)\left\| {\mathbf{v}_{kl}^I} \right\|_2^2}  \le {P_{BS}},
\end{equation}
where ${P_{BS}}$ is the maximum power each BS can provide. This power constraint model still has flaws that ${\left| {\mathbf{v}_{kl}^E} \right|^2}$ could be infinite when $\tau  \to 0$, and vise versa for ${\left| {\mathbf{v}_{kl}^I} \right|^2}$ in the case of $\tau  \to 1$. So the following additional constraints must be satisfied at the same time \cite{2-Nasir2017Beamforming}\cite{8-Zhang2013MIMO}.
\begin{equation}
\begin{array}{l}
\sum\limits_{k \in \mathcal{K}} {\left\| {\mathbf{v}_{kl}^E} \right\|_2^2}  \le {P_{peak}},\\
\sum\limits_{k \in \mathcal{K}} {\left\| {\mathbf{v}_{kl}^I} \right\|_2^2}  \le {P_{peak}},
\end{array}
\end{equation}
where ${P_{peak}} \ge {P_{BS}}$ is the maximum instantaneous transmit power of BS.

\subsection{Problem formulation}
To maximize the sum rate of the system, we formulate the joint design of BS transmit beamformer and the transmit TS ratio as an optimization problem (9), with the aforementioned design considerations (6)-(8) as constraints.
\begin{equation}
\mathop {\max }\limits_{{\mathbf{v}^I},{\mathbf{v}^E},\tau } {\rm{   }}\sum\limits_{k \in \mathcal{K}} {{R_k}} {\rm{ }}\quad\quad s.t.{\rm{  (6),(7),(8)}}.
\end{equation}
\par
Here, the conventional weighted sum rate(WSR) maximization problem is a well-known nonconvex optimization problem\cite{9-Dai2013Sparse}, for which finding the global optimal solution is quite challenging even without the chance energy harvesting constraint (6). To tackle the coupled relation of $\tau $ and ${{\mathbf{v}^I},{\mathbf{v}^E}}$, in this paper, we propose an AO algorithm to fetch a local optimum solution to the problem (9). 

\section{Alternative optimization algorithm for the sum rate maximization problem}

In this section, we first handle the chance constraint in (6) by Bernstein inequality, and utilize the AO algorithm to decouple the original problems as two sub-problems and settle them separately. 
\subsection{Bernstein approximation to the chance constraint of energy harvesting}

Because of the estimation error of CSI, it's difficult to guarentee the energy harvested by users with certainty. In this paper, a more practical method named chance constraint is adopted. Since there is no closed form solution to the chance constrained form in (6)\cite{13-Shen2012Distributed}, it is difficult to use convex optimization tools to solve the problem effectively. To solve this problem, the Bernstein inequality is adopted to obtain a conservative convex approximation. Next, the expression in (5) that the user's harvested energy with CSI error model is slightly deformed as
\begin{small}
\begin{equation}
\begin{array}{l}
{E_k} = \tau {\eta _k}\left( {\sum\limits_{j \in \mathcal{K}} {{{\left| {\mathbf{h}_k^H\mathbf{v}_j^E} \right|}^2}}  + {\sigma _k^2}} \right)\\
 = \tau {\eta _k}\left( {\sum\limits_{j \in \mathcal{K}} {{{\left| {\left( {\widehat{ \mathbf{h}}_k^H + \mathbf{e}_k^H} \right)\mathbf{v}_j^E} \right|}^2}}  + {\sigma _k^2}} \right)\\
 = \tau {\eta _k}\left[ {\sum\limits_{j \in \mathcal{K}} {\left( {\widehat{ \mathbf{h}}_k^H + \mathbf{e}_k^H} \right)\mathbf{v}_j^E{{\left( {\mathbf{v}_j^E} \right)}^H}\left( {{{\widehat{ \mathbf{h}}}_k} + {\mathbf{e}_k}} \right)}  + {\sigma _k^2}} \right]
\end{array}.
\end{equation}
\end{small}
In order to Adopt the Bernstein inequality, the chance harvested energy constraint of each user is expressed as,
\begin{equation}
{\rm{Pr}}\left\{ {{\mathbf{c}^H}{\mathbf{D}_k}\mathbf{c} + 2{\mathop{\rm Re}\nolimits} \left\{ {{\mathbf{c}^H}{\mathbf{d}_k}} \right\} \ge {\xi _k}} \right\} \ge 1 - \theta, 
\end{equation}
where ${\mathop{\rm Re}\nolimits} \{ \}$ represents the real part of the associated argument, and ${\mathbf{D}_k}$, ${\mathbf{d}_k}$, ${\xi _k}$ are defined as follows:
\begin{equation}
\begin{array}{l}
{\mathbf{D}_k} = \varepsilon _k^{{1 \mathord{\left/
 {\vphantom {1 2}} \right.
 \kern-\nulldelimiterspace} 2}}\sum\limits_{j \in \mathcal{K}} {\mathbf{v}_j^E{{\left( {\mathbf{v}_j^E} \right)}^H}} \varepsilon _k^{{1 \mathord{\left/
 {\vphantom {1 2}} \right.
 \kern-\nulldelimiterspace} 2}},\\
{\mathbf{d}_k} = \sum\limits_{j \in \mathcal{K}} {\varepsilon _k^{{1 \mathord{\left/
 {\vphantom {1 2}} \right.
 \kern-\nulldelimiterspace} 2}}\mathbf{v}_j^E{{\left( {\mathbf{v}_j^E} \right)}^H}{{\widehat{ \mathbf{h}}}_k}}, \\
{\xi _k} = \frac{{E_k^{\min }}}{{\tau {\eta _k}}} - \sigma _k^2 - \sum\limits_{j \in \mathcal{K}} {\widehat{ \mathbf{h}}_k^H\mathbf{v}_j^E{{\left( {\mathbf{v}_j^E} \right)}^H}{{\widehat{ \mathbf{h}}}_k}}. 
\end{array}
\end{equation}

Next, we will convert the nonconvex expression in (11) to a deterministic convex constraint based on Bernstein-type inequality\cite{14-Teng2017Robust}\cite{15-Bechar2009A}. Let $\mathbf{M} = {\mathbf{c}^H}\mathbf{Rc} + 2{\mathop{\rm Re}\nolimits} \left\{ {{\mathbf{c}^H}\mathbf{P}} \right\}$, where $\mathbf{R} \in {H^N}$ is a complex Hermitian matrix, $\mathbf{P} \in {C^N}$, $\mathbf{c} \sim \mathcal{CN} \left( {0,\mathbf{I}} \right)$, for any $\delta  > 0$,
\begin{equation}
{\rm{Pr}}\left\{ {\mathbf{M} \ge Tr(\mathbf{R}) - \sqrt {2\delta } \sqrt {\left\| \mathbf{R} \right\|_F^2 + 2{{\left\| \mathbf{P} \right\|}^2}}  - \delta {s^ - }} \right\} \ge 1 - {e^{ - \delta }}.
\end{equation}
\par
Let $\delta =  - \ln \theta $, (11) can be satisfied as long as the following inequation holds,
\begin{equation}
Tr\left( {{\mathbf{D}_k}} \right) - \sqrt {2\delta} \sqrt {\left\| {{\mathbf{D}_k}} \right\|_F^2 + 2{{\left\| {{\mathbf{d}_k}} \right\|}^2}}  - \delta s_k^ -  \ge {\xi _k},
\end{equation}
where $s_k^ -  = \sup \left\{ {\sup \left\{ {\lambda \left( { - {\mathbf{D}_k}} \right)} \right\},0} \right\}$, $\sup \left\{ {\lambda \left( { - {\mathbf{D}_k}} \right)} \right\}$ denotes the maximum eigenvalue of matrix $ - \mathbf{R}$. (14) can be transformed into three inequalities equivalently.
\begin{equation}
\begin{array}{l}
Tr\left( {{\mathbf{D}_k}} \right) - \sqrt {2{\delta}} x_k^E - {\delta}m_k^E \ge {\xi _k},\\
x_k^E \ge \sqrt {\left\| {{\mathbf{D}_k}} \right\|_F^2 + 2{{\left\| {{\mathbf{d}_k}} \right\|}^2}} ,\\
m_k^E\mathbf{I} + {\mathbf{D}_k} \succ 0,\\
m_k^E \ge 0.
\end{array}
\end{equation}

\subsection{Nonconvex problem decomposition}

Due to that ${\mathbf{D}_k}$, ${\mathbf{d}_k}$ , ${\xi _k}$ are indefinite quadratic in $\mathbf{v}_j^E$, we adopt SDR technology to solve the problem efficiently. Let $\mathbf{V}_j^E = \mathbf{v}_j^E{\left( {\mathbf{v}_j^E} \right)^H}$ be a semi-definite symmetric matrix and $rank\left( {\mathbf{V}_j^E} \right) = 1$. Furthermore, the following equations holds.
\[\left\| {\mathbf{v}_{kl}^E} \right\|_2^2 = Tr\left[ {{\mathbf{A}_l}\mathbf{V}_k^E{{\left( {{\mathbf{A}_l}} \right)}^H}} \right], \quad\quad \left\| {\mathbf{v}_{kl}^I} \right\|_2^2 = \left\| {{\mathbf{A}_l}\mathbf{v}_k^I} \right\|_F^2.\]
where ${\mathbf{A}_l} \in {C^{N \times N}}$ is a block diagonal matrix in which the diagonal element of the $l$ diagonal block is 1, and the remainder is 0. The power constraint of each BS in (7) can be transformed into
\begin{small}
\begin{equation}
\sum\limits_{k \in \mathcal{K}} {\tau Tr\left[ {{\mathbf{A}_l}\mathbf{V}_k^E{{\left( {{\mathbf{A}_l}} \right)}^H}} \right]}  + \sum\limits_{k \in \mathcal{K}} {\left( {1 - \tau } \right)\left\| {{\mathbf{A}_l}\mathbf{v}_k^I} \right\|_F^2}  \le {P_{BS}}.
\end{equation}
\end{small}
Then, the optimization problem (9) can be equivalently transformed to

\begin{subequations}

\begin{align}
\mathop {\max }\limits_{{\mathbf{v}^I},{\mathbf{V}^E},\tau } &{\rm{   }}(1 - \tau )\sum\limits_{k \in \mathcal{K}} {\log \left( {1 + SIN{R_k}} \right)} \\
\quad s.t. &{\rm{     }}\sum\limits_{k \in \mathcal{K}} {Tr\left[ {{\mathbf{A}_l}\mathbf{V}_k^E{{\left( {{\mathbf{A}_l}} \right)}^H}} \right]}  \le {P_{peak}}\\
 &{\rm{}}\sum\limits_{k \in \mathcal{K}} {\left\| {{\mathbf{A}_l}\mathbf{v}_k^I} \right\|_F^2}  \le {P_{peak}}{\rm{      }}\\
\quad\quad\quad\quad  &{\rm{}}(15),(16)
\end{align}
\end{subequations}

However, this problem is nonconvex about ${{\mathbf{v}^I},{\mathbf{V}^E},\tau }$ simultaneously, but fortunately, when $\{{\mathbf{v}^I},{\mathbf{V}^E}\}$ or $\tau $ is fixed, the remaining problem can be transformed to the strictly convex form. In this paper, an AO algorithm is used to solve this nonconvex problem. Two sub-problems with respect to ${\mathbf{v}^I},{\mathbf{V}^E}$ and $\tau $ are given as below,
\\
\textbf{Problem 1: Joint solution of information transmission beamformer and energy transmission beamformer}

\[\begin{array}{l}
\mathop {\max }\limits_{{\mathbf{v}^I},{\mathbf{V}^E}} {\rm{   }}(1 - \tau )\sum\limits_{k \in \mathcal{K}} {\log \left( {1 + SIN{R_k}} \right)} \\
\quad {\rm{ }}s.t. \quad {\rm{    (17b),(17c),(17d).}}
\end{array}\]
\\
\textbf{Problem 2: Solution for time ratio of SWIPT}

\[\begin{array}{l}
\mathop {\max\quad } \limits_\tau  (1 - \tau )\sum\limits_{k \in \mathcal{K}} {{R_k}} \\
\quad s.t. \quad 0 < \tau  < 1,\\
\qquad\quad (17d).
\end{array}\]

Note that $\mathbf{v}^I$ and $\mathbf{V}^E$ in problem 1 is a typical MSE problem, we consult to its equivalent MMSE solution\cite{9-Dai2013Sparse}\cite{10-Shi2011An}. As to problem 2, since it can be transformed to the linear optimization form in (18), it can be easily solved through CVX directly.

\begin{equation}
\begin{array}{l}
\mathop {\min }\limits_\tau  {\rm{   }}\tau \sum\limits_{k \in \mathcal{K}} {{R_k}} {\rm{}} \quad
s.t. \quad {\rm{       }}0 < \tau  < 1,
{\rm{       (17d).}}
\end{array}
\end{equation}

\subsection{MMSE solution for Problem 1}

In view of the WSR problem, we assume ${\mathbf{w}_k}$ is the receiver pre-coding, and the mean square error is defined as
\begin{equation}
\begin{array}{l}
{M_k} = E\left\{ {\left( {\mathbf{w}_k^H{y_k} - {s_k}} \right){{\left( {\mathbf{w}_k^H{y_k} - {s_k}} \right)}^H}} \right\}\\
\quad {\rm{     }} = \mathbf{w}_k^H\left( {\sum\limits_{j \in \mathcal{K}} {\mathbf{h}_k^H\mathbf{v}_j^I{{\left( {\mathbf{v}_j^I} \right)}^H}{\mathbf{h}_k}}  + \sigma _k^2\mathbf{I}} \right){\mathbf{w}_k} \\
\quad\quad - 2{\mathop{\rm Re}\nolimits} \left\{ {\mathbf{w}_k^H\mathbf{h}_k^H\mathbf{v}_k^I} \right\} + 1.
\end{array}
\end{equation}
\par
For the imperfect CSI, the average mean square error is considered \cite{11-Fritzsche2013Robust}, which results in 
\begin{equation}
\begin{array}{l}
{{\bar M}_k} = {{E_\mathbf{e_k}}}\left( {{M_k}} \right) = 1 + \sigma _k^2\mathbf{w}_k^H{\mathbf{w}_k} 
- 2{\mathop{\rm Re}\nolimits} \left\{ {\mathbf{w}_k^H\widehat{ \mathbf{h}}_k^H\mathbf{v}_k^I} \right\}
{\rm{}}\\
+ {E_\mathbf{e_k}}\left\{ {\sum\limits_{j \in \mathcal{K}} {\mathbf{w}_k^H\mathbf{e}_k^H\mathbf{v}_j^I{{\left( {\mathbf{v}_j^I} \right)}^H}{\mathbf{e}_k}{\mathbf{w}_k}} } \right\}\\
+ \sum\limits_{j \in \mathcal{K}} {\mathbf{w}_k^H\widehat{ \mathbf{h}}_k^H\mathbf{v}_j^I{{\left( {\mathbf{v}_j^I} \right)}^H}{{\widehat{ \mathbf{h}}}_k}{\mathbf{w}_k}}.
\end{array}
\end{equation}

Taking expectation of (20) with respect to $\mathbf{e}_k$, we consider the lower bound of user rate\cite{11-Fritzsche2013Robust} :
\begin{equation}
{E_{\mathbf{{e_k}}}}\left\{ { - \log \det \left( {{M_k}} \right)} \right\} \ge  - \log \det \left( {{E_\mathbf{e_k}}\left\{ {{M_k}} \right\}} \right) = {\bar R_k}.
\end{equation}
\par
With (21), we can transform the original problem into maximizing the lower bound of average sum rate. Besides, the expectation in (20) can be calculated as
\begin{equation}
{E_{\mathbf{{e_k}}}} \left\{ {\sum\limits_{j \in \mathcal{K}} {\mathbf{w}_k^H\mathbf{e}_k^H\mathbf{v}_j^I{{\left( {\mathbf{v}_j^I} \right)}^H}{\mathbf{e}_k}{\mathbf{w}_k}} } \right\} = \mathbf{w}_k^H{\Phi _k}{\mathbf{w}_k},
\end{equation}
where ${\Phi _k} = Tr\left( {\sum\limits_{j \in \mathcal{K}} {\mathbf{v}_j^I{{\left( {\mathbf{v}_j^I} \right)}^H}} } \right)*{\varepsilon _k} = Tr\left( {{{\left( {{\mathbf{v}^I}} \right)}^H}{\mathbf{v}^I}} \right)*{\varepsilon _k} = \left\| {{\mathbf{v}^I}} \right\|_F^2*{\varepsilon _k}$. Now, we can solve problem 1 by the block coordinate descent(BCD) method \cite{10-Shi2011An}, which is implemented through  four steps as bellow.
\\
\textbf{Step 1:} Minimize the sum MSE leads to the MMSE receiver:
\begin{equation}
\mathbf{w}_k^{MMSE} = \frac{{\mathbf{h}_k^H\mathbf{v}_k^I}}{{\sum\limits_{j \in \mathcal{K}} {\mathbf{h}_k^H\mathbf{v}_j^I{{\left( {\mathbf{v}_j^I} \right)}^H}{\mathbf{h}_k}}  + \sigma _k^2\mathbf{I} + {\Phi _k}}}.
\end{equation}
\textbf{Step 2:} Under the MMSE receiver, the corresponding MSE is expressed as:
\begin{equation}
\bar M_k^{MMSE} = 1 - \frac{{{{\left( {\mathbf{v}_k^I} \right)}^H}{{\widehat{ \mathbf{h}}}_k}\widehat{ \mathbf{h}}_k^H\mathbf{v}_k^I}}{{\sum\limits_{j \in \mathcal{K}} {\mathbf{h}_k^H\mathbf{v}_j^I{{\left( {\mathbf{v}_j^I} \right)}^H}{\mathbf{h}_k}}  + \sigma _k^2I + {\Phi _k}}}.
\end{equation}
\textbf{Step 3:} According to the mean square error calculated by (24), we can get the weight matrix
\begin{equation}
{\rho _k} = \bar M_k^{ - 1}.
\end{equation}
\textbf{Step 4:} Fixing $\mathbf{w}_k^{MMSE}$ and ${\rho _k}$, the optimal solution of problem 1 is equivalent to the minimization problem as follows
\begin{equation}
\begin{array}{l}
\mathop {\min }\limits_{{\mathbf{v}^I},{\mathbf{V}^E}} {\rm{   }}(1 - \tau )\sum\limits_{k \in \mathcal{K}} {\left( {{\rho _k}{{\bar M}_K} - \log \det {\rho _k}} \right)} \\
\;\; s.t. \quad (17b),(17c),(17d).
\end{array}
\end{equation}
\par
Here, $\rho _k$ is calculated from the previous iteration. Replacing the objective function in (26) by (20), and removing the items that does not affect the result, the object in (26) can be further equivalent to

\begin{equation}
\sum\limits_{k \in \mathcal{K}} {{\rho _k}\left( \begin{array}{l}
\sum\limits_{j \in \mathcal{K}} {\mathbf{w}_k^H\widehat{ \mathbf{h}}_k^H\mathbf{v}_j^I{{\left( {\mathbf{v}_j^I} \right)}^H}{{\widehat{ \mathbf{h}}}_k}{\mathbf{w}_k}}  - 2{\mathop{\rm Re}\nolimits} \left\{ {\mathbf{w}_k^H\widehat{ \mathbf{h}}_k^H\mathbf{v}_k^I} \right\}\\
 + {\varepsilon _k}*\mathbf{w}_k^HTr\left( {{{\left( {{\mathbf{v}^I}} \right)}^H}{\mathbf{v}^I}} \right){\mathbf{w}_k}
\end{array} \right)}.
\end{equation}
\par
This problem is already a convex optimization problem and can be solved by CVX effectively\cite{12-Grant2008CVX}. Note that problem (26) is a conservative approximation of problem (9) since that the optimal solution of (26) may not be rank one. When the solution of (26) satisfies the rank one condition for all users, we can obtain $\mathbf{V}_j^E = \mathbf{v}_j^E{\left( {\mathbf{v}_j^E} \right)^H}$ , and $\mathbf{v}_j^E$ is a feasible and optimal solution to problem (26). However, if the solution does not satisfy the rank one condition, some additional approximate procedure is needed to transform the solution to rank one, such as Gaussian Randomization. But fortunately, with the conclusion of \cite{13-Shen2012Distributed}, the optimization solution $\mathbf{V}_j^E$ is always rank one, and furthermore, the results in our simulation also consistent with the conclusion. Therefore, applying SDR to the problem does not increase the conservativeness of the solution.

Furthermore, an AO algorithm is adopted to get a local optimum solution of (17). The algorithm consists of two loops: inner loop and outer loop, where the outer loop is mainly to solve Problem 1 and Problem 2 alternately, while the inner loop is used to solve the approximate WSR problem. The detail of AO algorithm is demonstrated in \textbf{Algorithm 1}.  

~\\
{\color{black}
\begin{tabular}{m{8cm}}\Xhline{1pt}
\textbf{Algorithm 1:} The AO Algorithm\\
\Xhline{1pt}
\textbf{Step 1:} Initialize iteration counter r=0, maximum number of iterations $r_{max}=20$ and $\tau  = 0.5$.\\

\textbf{Step 2 :} Solve problem 1:\\
\textbf{Repeat:} \\
1) Calculate the MMSE receiver $\mathbf{w}_k^{MMSE}$ and the corresponding average MMSE $\bar M_k^{MMSE}$ according to (23) and (24);\\
2) Update the MSE weight ${\rho _k}$ according to (25);\\
3) Find the optimal beamformer ${\mathbf{v}^I}$ and ${\mathbf{V}^E}$ by (26) under fixed  and $\mathbf{w}_k^{MMSE}$.\\
\textbf{Until convergence.}\\

\textbf{Step 3 :} Solve problem 2 under the result of Step 2.\\

\textbf{Step 4 :} \textbf{If} $r = r_{max}$ or converge, obtain the optimal beamforming vector.\\
\textbf{Else} go to \textbf{Step 2}\\

\textbf{Step 5 :} Get beamforming vector $\mathbf{v}_k^{I[r]},\mathbf{v}_k^{E[r]},\forall k \in \mathcal{K}$ from the solution $\mathbf{V}_k^{I[r]},\mathbf{V}_k^{E[r]}$.\\
\textbf{End}\\
\Xhline{1pt}
\end{tabular}
~\\
}

\section{Simulation and analysis}

In this section, numerical results are provided to demonstrate the effectiveness of the proposed algorithm. In this simulation, a network consisting of 5 BSs and 3 users is considered, and each BS has 2 antennas, each user has only one antenna. The BSs and users are distributed in a region of 800*800 independently, and the distribution of the user and the BSs follow the Homogeneous Poisson Point Process (HPPP). Taking into account large scale fading and small-scale fading, the channel parameters is presented in TABLE 1\cite{14-Teng2017Robust}. In addition, we assume that the CSI error follows the independent complex Gauss distribution. Finally, we assume that all users are consistent in terms of energy requirements, and that the maximum outage probability of energy harvesting user can tolerate is 0.2. Besides, energy conversion efficiency of energy harvesting is 0.75. The rest of simulation parameters are listed in TABLE 1. 

\begin{center}
		\footnotesize
		\vspace{-0.1em}
		TABLE I. 	SIMULATION PARAMETERS\par~\\
		\begin{tabular}{|p{140pt}|p{75pt}|}
			\hline
			\begin{minipage}{120pt}\centering Simulation Parameter \end{minipage} &
			\begin{minipage}{80pt}\centering Value \end{minipage} \\
			\hline
			
			Path-loss at distance d (km) & 128.1 + 37.6*log10(d)\\
			\hline
			Transmit antenna power gain & \quad\quad\quad\quad9dBi\\
			\hline
			Standard deviation of log-norm shadowing & \quad\quad\quad\quad6dB\\
			\hline
			Small-scale fading distribution & \qquad\quad $ {\cal C}{\cal N}(0,{\mathbf{I}_N})$\\
			\hline
			Noise variance of all users & \quad\quad$\sigma _k^2 = 0.01$\\
			\hline
			Maximum transmission power of BS &\quad\quad ${P^{BS}} = 5W$\\
			\hline
			Variance of CSI error distribution & \qquad\quad ${\varepsilon _k} = 0.01$\\
			\hline
			Maximum instantaneous transmission power of BS &\quad\quad ${P^{peak}} = 5W$\\
			\hline

		\end{tabular}		
\end{center}

Fig.2 shows how the sum rate varies with the number of iterations under different energy outage probabilities. As can be seen from the diagram, the higher the energy outage probability is, the greater sum rate the system can achieve. Meanwhile, the algorithm with deterministic constraint condition can achieve the lowest rate. This is consistent with the actual meaning, because the lower outage probability represents the system constraints on the harvested energy are more stricter, and it will take more resources to transfer energy to the user which leads to a relative reduction of energy for information transmission. Furthermore, it can be seen from Fig.2 that as the number of iterations increases, the sum rate gradually increased and eventually stabilized, and this phenomenon shows that the AO algorithm proposed in this paper tends to be optimal through iterations.

\begin{center}
\includegraphics[width=3.3in]{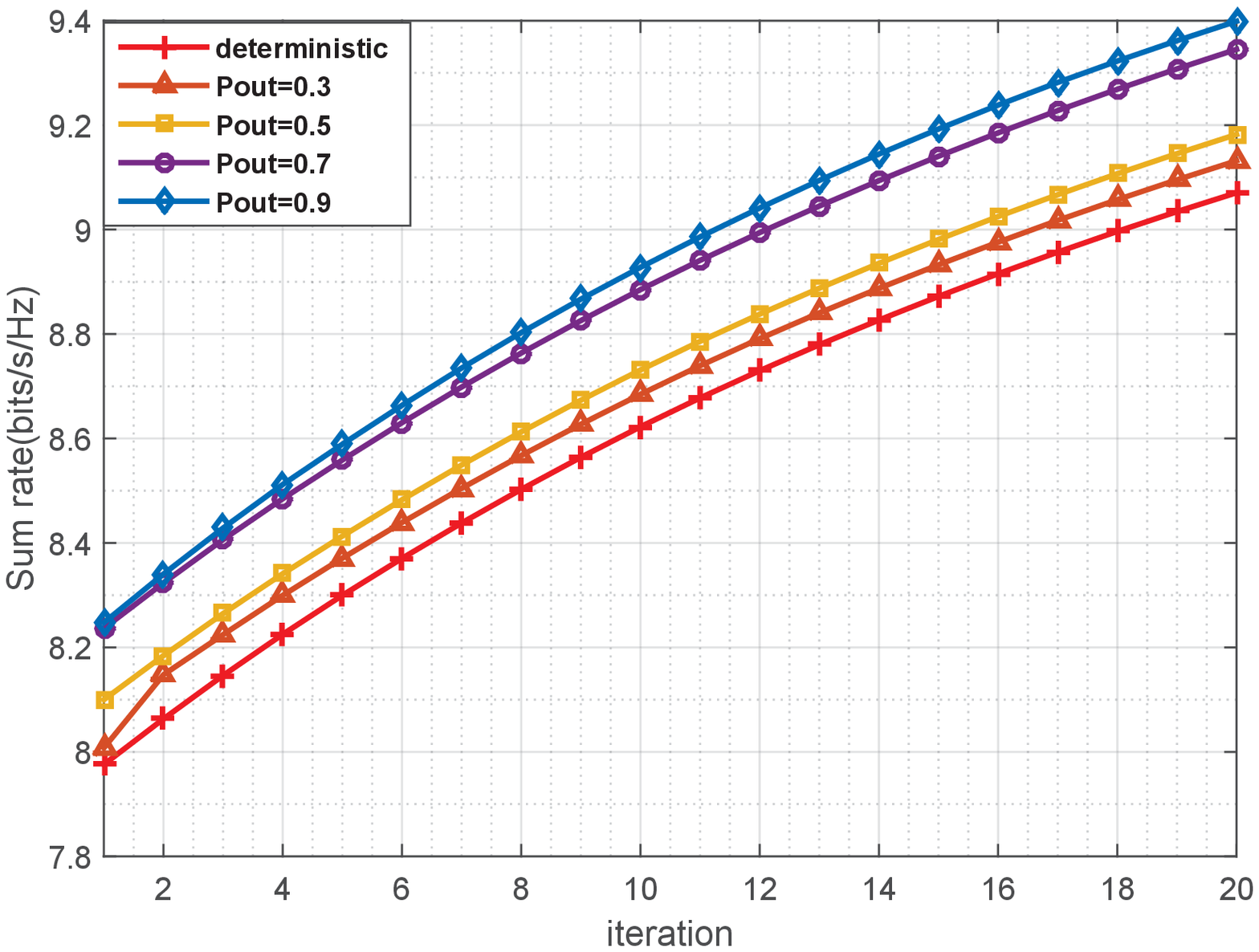}
\\
\footnotesize
Fig.2 Maximum sum rate evolutions of iterations
\end{center}

\begin{center}
\includegraphics[width=3.3in]{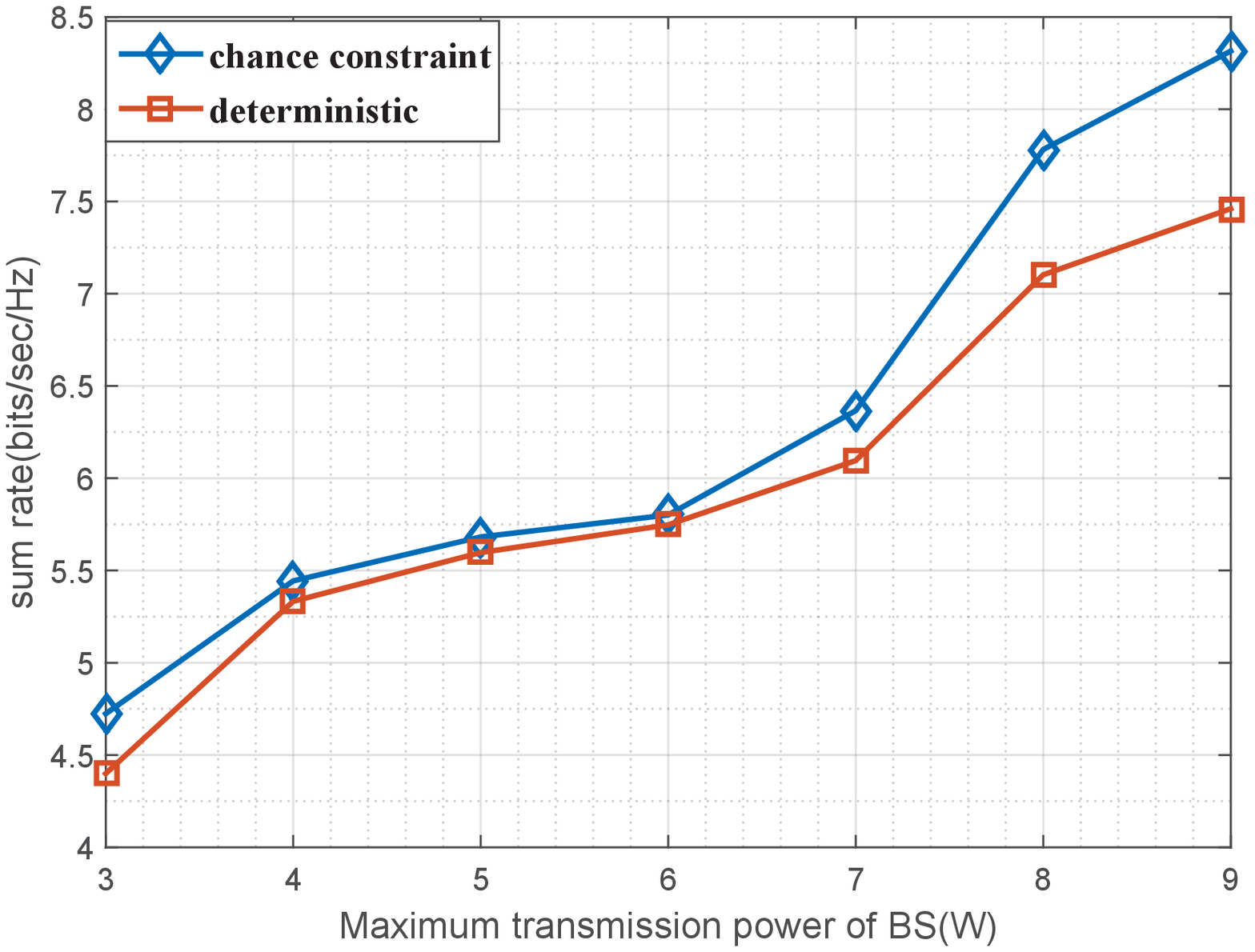}\\
\footnotesize
Fig.3 Maximum sum rates vary with the maximum transmission power
\end{center}

Fig.3 illustrates the relationship between the sum rate and the maximum transmission power of the BS under the chance constraint and the deterministic constraint respectively. From the graph, we can find that although the two curves are closer in some cases, the sum rate of chance constrained algorithms are higher than those using deterministic constraints. Moreover, with the increase of the maximum transmission power of the BS, the sum rate is increasing, and this phenomenon indicates that under the premise of satisfying the user's energy collection, the base station will use as much energy as possible for the information transmission, so that users can have higher transmission rates.
\\
\begin{center}
\footnotesize
		\vspace{-0.1em}
		TABLE II. 	Optimzed TS ratio $\tau$ for different minimum EH threshold\par~\\
\begin{tabular}{|l|c|c|c|c|c|} 
\hline 
EH threshold&0.05&0.1&0.15&0.2&0.25\\
\hline  
TS ratio $\tau$&0.548&0.551&0.558&0.561&0.565\\
\hline 
\end{tabular}
\end{center}

\begin{center}
\includegraphics[width=3.3in]{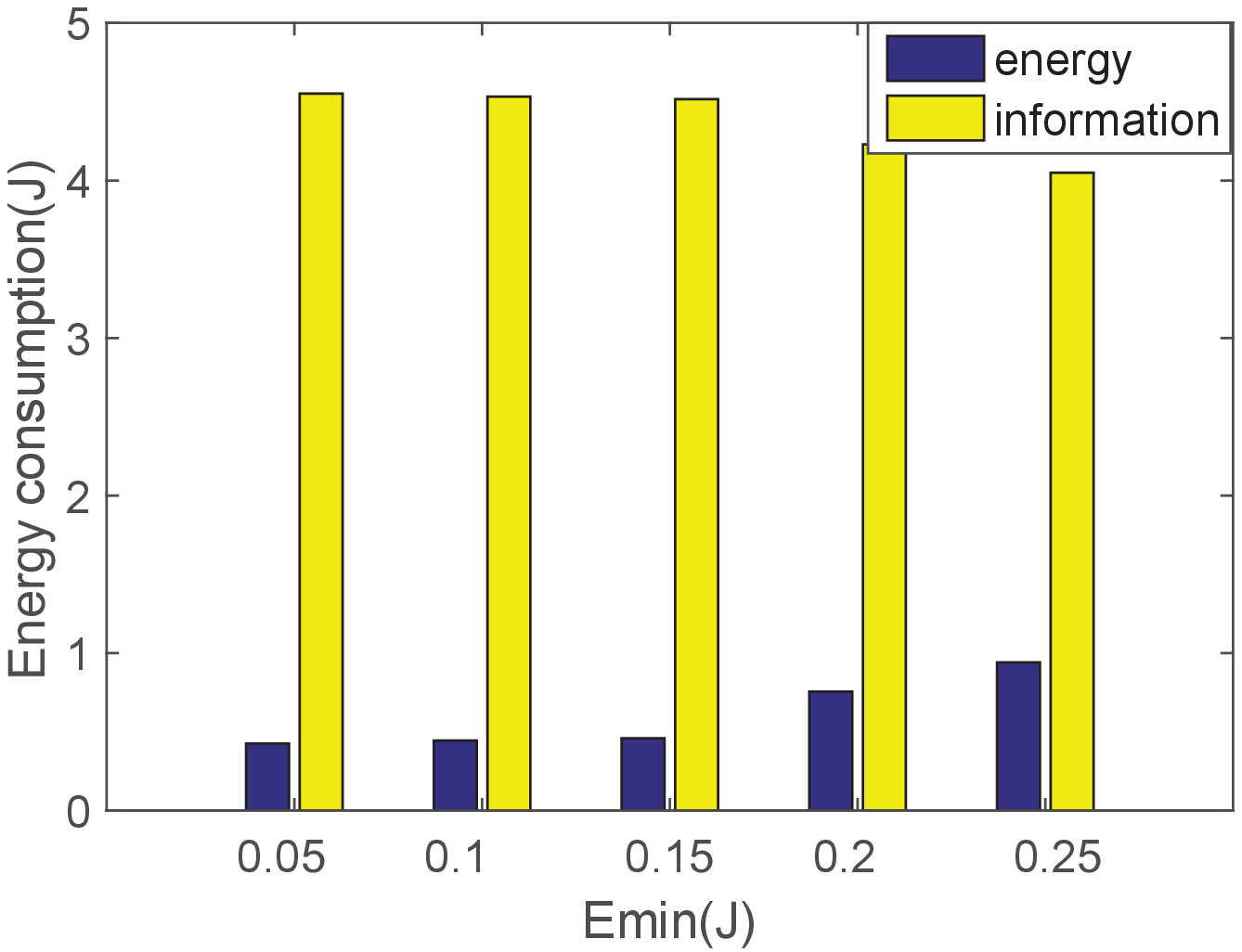}\\
\footnotesize
Fig.4 Energy consumption for transmitting information and energy
\end{center}

Assuming that the maximum energy consumption BS can provide is 5 (Joule), the energy consumption used to transmit energy and transmit information is shown in Fig.4, and the corresponding TS ratio is shown in table II. As can be seen, the energy consumption of the BS for energy transmission and the TS ratio increase with the increase of the requested minimal energy, while the the energy consumption for information transmission decreases correspondingly. 

\begin{center}
\includegraphics[width=3.3in]{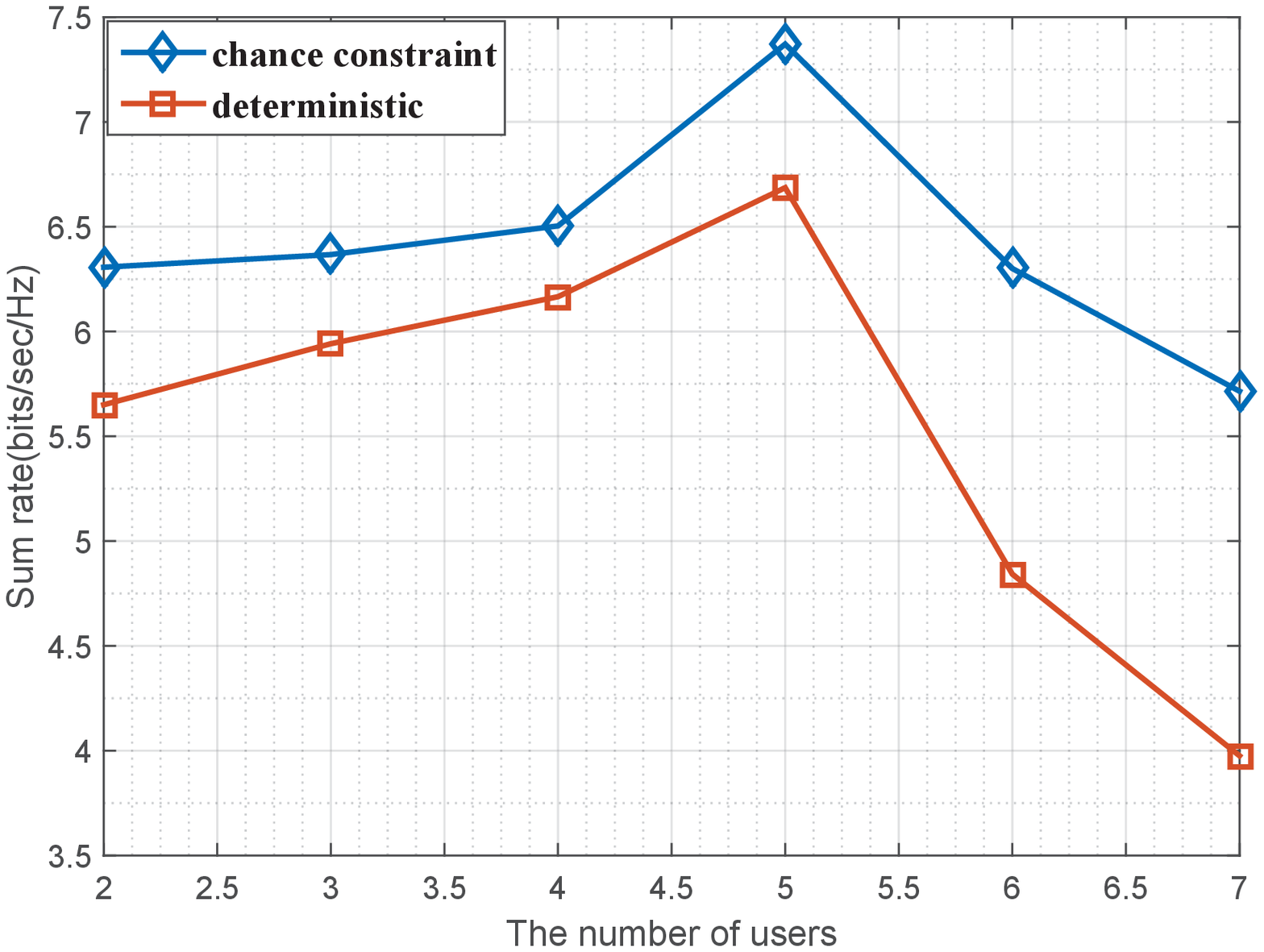}\\
\footnotesize
Fig.5 Maximum sum rate for different user density
\end{center}

In Fig.5, with the number of users varying from 2 to 7. It is seen that the sum rate taking the first growing and then decreasing trend, and the reason for this phenomenon is that the BSs have potential capacity to provide higher sum rate to these users when the number of users is rather small. However, when the number of users increases to a threshold, more energy is cost for each user's energy harvesting, and this will inevitably leads to the reduction of power consumption for information transmission, and the sum rate reduces accordingly. In addition, as a whole can be seen in the experiment, the sum rate of chance constrained beamforming algorithm is always higher than the deterministic constrained scene. It also indicates that the chance constraint beamforming algorithm can provide higher system rates under the premise of ensuring the normal operation of the user.  

\section{Conclusions}

In this paper, we present a formulation of robust beamforming with objective of maximizing the sum rate in SWIPT network, and proposed an AO algorithm to solve the problem effectively. The chance constraints for energy harvesting are transformed into convex deterministic constraints, and MMSE method is employed to solve the maximum sum rata problem equivalently. Simulation results have demonstrated the advantages of proposed algorithm to deterministic ones.


\end{document}